\documentclass[letter, 11pt]{article}

\usepackage{fullpage}
\usepackage{psfrag}
\usepackage{epsfig}
\usepackage{tabularx}
\usepackage{multirow}
\usepackage{subfigure}
\usepackage{multicol}
\usepackage{color}
\usepackage[active]{srcltx}
\usepackage{amssymb}
\usepackage{amsmath,wasysym}
\usepackage{color}
\usepackage{cite}
\allowdisplaybreaks[2]          
\usepackage{amssymb} 
\usepackage{amsmath}
\allowdisplaybreaks[2]          
\usepackage[figure,noend]{algorithm}
\usepackage{graphicx}
\usepackage{subfigure}
\usepackage{url}
\usepackage{xspace}
\usepackage{times}
\newcommand{\circled}[1]{\raisebox{.5pt}{\textcircled{\raisebox{-.9pt} {#1}}}}

\usepackage{float}
\floatstyle{ruled}
\newfloat{alg}{htbp}{alg}
\floatname{alg}{Algorithm}

\usepackage[noend]{distribalgo}
\newfloat{algo}{htbp}{algo}
\floatname{algo}{Algorithm}

\newcommand{\omitit}[1]{}

\usepackage[active]{srcltx}
\usepackage{amssymb} 

\newcommand{\N}{\mathbb{N}}

\newcommand{\protocol}{\emph{MDStore}\xspace}

\newtheorem{definition}{Definition}[section]
\newtheorem{lemma}[definition]{Lemma}

\newtheorem{theorem}[definition]{Theorem}

\newcommand*{\qed}{\hfill\ensuremath{\square}}%
\newenvironment{prooff}{\vspace{1ex}\noindent{\bf Proof:}\hspace{0.5em}}
	{\hfill\qed\vspace{1em}}

\title{Asynchronous BFT Storage with $2t+1$ Data Replicas}
\author{
  Christian Cachin\\
  IBM Research - Zurich\\
  \texttt{cca@zurich.ibm.com}
  \and
  Dan Dobre\\
  NEC Labs Europe\\
  \texttt{dan.dobre@neclab.eu}
  \and
  Marko Vukoli\'{c}\\
  EURECOM\\
  \texttt{marko.vukolic@eurecom.fr}
}

\date{}
\begin{document}

\maketitle
\thispagestyle{empty}

\begin{abstract}
The cost of Byzantine Fault Tolerant (BFT) storage is the main concern preventing its adoption in practice. This cost stems from the need to maintain at least $3t+1$ replicas in different storage servers in the asynchronous model, so that $t$ Byzantine replica faults can be tolerated.

In this paper, we present \protocol, the first fully asynchronous read/write BFT storage protocol that reduces the number of data replicas to as few as $2t+1$, maintaining $3t+1$ replicas of metadata at (possibly) different servers. At the heart of \protocol store is its metadata service that
is built upon a new abstraction we call \emph{timestamped storage}. Timestamped storage both allows for conditional writes (facilitating the implementation of a metadata service) and has consensus number one (making it implementable wait-free in an asynchronous system despite faults). In addition to its low data replication factor, \protocol offers very strong guarantees implementing multi-writer multi-reader atomic wait-free semantics and tolerating any number of Byzantine readers and crash-faulty writers.

We further show that \protocol data replication overhead is optimal; namely, we prove a lower bound of $2t+1$ on the number of data replicas that applies even to crash-tolerant storage with a fault-free metadata service oracle.
Finally, we prove that separating data from metadata for reducing the cost of BFT storage is not possible without cryptographic assumptions. However, our \protocol protocol uses only lightweight cryptographic hash functions.
\end{abstract}



\newpage
\setcounter{page}{1}

\section{Introduction}
\label{sec:intro}

\newcommand{\td}{t}
\newcommand{\bd}{b}
\newcommand{\tm}{t_M}

Byzantine Fault Tolerant (BFT) protocols are notoriously costly to deploy. This cost stems from the fact that, in many applications,  tolerating Byzantine faults requires more resources than tolerating less severe faults, such as crashes. For example, in the asynchronous communication model,  BFT read/write storage protocols~\cite{Lam86} are shown to require at least $3t+1$ replicas in different storage servers so that $t$ Byzantine server faults can be tolerated~\cite{MAD02}. This is to be contrasted with the requirement for $2t+1$ replicas in the asynchronous crash model for protocols used in production cloud-storage systems. This gap between crash tolerance and BFT is one of the main concerns for practical adoption of BFT systems.

In this paper we show that this gap may in fact be significantly smaller. Namely, we present
\protocol, a novel asynchronous message-passing read/write storage emulation that reduces the number of \emph{data replicas} to only $2\td+1$, maintaining $3\tm+1$ \emph{metadata replicas} at (possibly) different servers. Here, $\td$ and $\tm$ are thresholds on the number of Byzantine data and metadata replicas, respectively. To achieve lower replication cost, \protocol does not sacrifice other functionalities. Namely, \protocol implements multi-writer multi-reader (MWMR) atomic wait-free storage~\cite{Lam86,Her91} that tolerates any number of Byzantine readers and crash-faulty writers. \protocol is the first asynchronous BFT storage protocol that does not assume any trusted components to reduce its resource cost (unlike~\cite{Correia:2004:THL,Chun:2007:AAM,KapitzaBCDKMSS12,VeroneseCBLV13}). Moreover, being a fully asynchronous read/write storage protocol, \protocol is fundamentally different from the existing consensus~\cite{GV10}, state-machine replication (SMR)~\cite{PaxosMadeSimple, Yin03} and SMR-based storage protocols~\cite{Farsite}, which employ the similar separation of control and data planes and which are all subject to the FLP impossibility result~\cite{FLP85} and require partial synchrony~\cite{DLS88}.

\protocol has modular architecture: a client reads and writes metadata (which consists of a hash of the value, timestamp and pointers to data replicas that store a value) through the abstraction of a \emph{metadata service}. The Metadata service is an array of SWMR safe wait-free storage objects~\cite{Lam86} and a novel MWMR atomic wait-free storage object variant, which we call \emph{timestamped storage}. In an array of safe storage, indexed by timestamps, \protocol stores hashes of data values, whereas in atomic timestamped storage, \protocol stores pointers to $\td+1$ (out of $2\td+1$) data replicas storing the most recent value. On the other hand, data replicas simply store timestamp/value pairs.

Our timestamped storage object is very similar to classical atomic~\cite{Lam86} (or linearizable~\cite{HW90}) read/write storage, except that it also exposes a timestamp attached to the stored values to clients, allowing for conditional writes, i.e., writes that take effect conditional on a timestamp value. Interestingly, despite its support of conditional writes, timestamped storage has consensus number~\cite{Her91} equal to one,  which makes an implementation of the metadata service possible in the asynchronous model despite faults. Indeed, we show that the \protocol metadata service can
be implemented from simple asynchronous BFT SWMR safe~\cite{MR98, ACKM06, GV06} and SWMR atomic~\cite{Phalanx,Cachin:2006:ORE,AAB07,PoW} storage protocols using $3t+1$ replicas for tolerating $t$ faults; in the context of \protocol, these replicas are exactly the $3\tm+1$ \emph{metadata replicas}.

Complementing the \protocol protocol, this paper also establishes lower bounds on the number of data replicas that are needed for asynchronous storage implementations with logically separated metadata.  In more detail:
\begin{itemize}
\item We prove that at least $2\td+1$ data replicas are necessary for implementations that leverage a metadata service, even if data replicas can fail only by crashing. This shows not only that \protocol is optimally resilient, but also that it incurs no additional data replication cost compared to crash-tolerant storage. The lower bound of $2\td+1$ has a very broad scope: it applies already to obstruction-free~\cite{Herlihy:2003:OSD} single-writer single-reader safe storage~\cite{Lam86} (and can be extended to eventual consistency~\cite{Vogels09}). Moreover, for the purpose of the lower bound, we define a metadata service very loosely as a fault-free oracle that provides arbitrary functionality with the single limitation that it cannot store or forward data values, roughly speaking. We believe that this definition of a metadata service is of independent interest.

\item We show that reducing the cost of BFT storage by separating metadata and data requires to limit the computational power of a Byzantine adversary. In the practically relevant case of a bounded adversary that cannot subvert collision resistance of cryptographic hash functions, \protocol shows that $2\td+1$ data replicas are sufficient. However, with an unbounded adversary, we show that one needs on $3\td+1$ data replicas, despite the metadata service oracle. 
\end{itemize}


The rest of the paper is organized as follows. In Section~\ref{sec:model} we introduce the system model and preliminary definitions. Section~\ref{sec:protocol} presents \protocol. In Section~\ref{sec:lb} we prove our lower bounds on the number of data replicas and Section~\ref{sec:relwork} discusses related work. Finally, Section~\ref{sec:conclusion} concludes the paper with an outlook to future work.  The correctness proof of \protocol is postponed to Appendix~\ref{app:proof}.

\section{System model and definitions}
\label{sec:model}


\noindent\textbf{Processes.} The distributed system we consider consists of four
sets of processes: (i) a set \emph{metadata replicas} of size $M$ containing
processes $\{m_1, ..., m_M\}$, (ii) a set of $D$ \emph{data replicas} containing
processes $\{d_1, ..., d_D\}$, (iii) a set of $W$ \emph{writers} containing
processes $\{w_1, ..., w_W \}$; and (iv) a set \emph{readers} of size $R$ containing processes
$\{r_1, ..., r_R \}$.
The set \emph{clients} is the union of \emph{writers} and \emph{readers}. Similarly, the set \emph{replicas} denotes the union of \emph{data replicas} and \emph{metadata replicas}. Clients are disjoint from replicas, but \emph{writers} and \emph{readers} may intersect, just like \emph{metadata replicas} and \emph{data replicas}b.  Clients are either \emph{benign} or \emph{Byzantine}, as defined later.

We model distributed algorithm $A$ for set of processes $\Pi$ as a collection of deterministic automata, where $A_p$ is the automata assigned to process $p\in\Pi$.  The computation of benign processes proceeds in \emph{steps} of $A$.  For space constraints, we omit the details of this model and refer to the literature~\cite{LT89}.

\noindent\textbf{Channels.} We assume that every process can communicate with every other process using
point-to-point perfect asynchronous communication channels \cite{CachinGR11}. In short, perfect channels guarantee reliable communication: i.e.,  if neither process at the end of a communication channel is faulty, every sent message is
eventually delivered to the receiver exactly once.\footnote{Perfect channels are simply implemented from lossy channels using retransmission mechanisms \cite{CachinGR11}.} For presentation simplicity, we also assume a global clock, which, however, is not accessible to processes who perform local computations and communicate asynchronously.


\noindent\textbf{Adversary.} A \emph{Byzantine} process $p$ does not follow $A_p$ and may perform arbitrary \emph{actions}, such as  (i) sending arbitrary messages or (ii) changing its state in an arbitrary manner. We assume an \emph{adversary} that can coordinate Byzantine processes and make them collude. 

We use a deterministic model for a cryptographic hash function.  A
hash function maps a bit string of arbitrary length to a short, unique
representation of fixed length and consists of a distributed oracle
accessible to all processes.  The hash oracle exports a single
operation $H$; its invocation takes a bit string $x$ as parameter and
returns an integer as the response.  The oracle maintains a list $L$
of all $x$ that have been invoked so far.  When the invocation
contains $x \in L$, then $H$ responds with the position of $x$ in $L$;
otherwise, $H$ appends $x$ to the end of $L$ and returns its position.
This ideal implementation models only collision resistance, i.e., that
it is infeasible even for an unbounded adversary to produce two
different inputs $x$ and $x'$ such that $H(x) = H(x')$.  

In the following, unless explicitly specified differently, we use this
model of a hash function.  In our context this is equivalent to
assuming that the adversary is computationally \emph{bounded}, i.e.,
that it cannot break cryptographic hash functions.  Alternatively, we speak
of an \emph{unbounded} adversary when no such hash function is
available.  This terminology matches the traditional names and
formalizations of cryptographic hash functions~\cite{KatzLindell07}.

Finally we assume that channels that relate benign processes are
\emph{authenticated}, i.e., that the adversary cannot (undetectably)
insert messages in these channels.  In practice, authenticated
communication can be implemented easily from point-to-point channels
with a mes\-sage-authen\-ti\-ca\-tion code (MAC)~\cite{KatzLindell07}.

\noindent\textbf{Executions and faults.} Given any algorithm $A$, an \emph{execution} of $A$ is an infinite sequence
of steps of $A$ taken by benign processes, and actions of
Byzantine processes.
A \emph{partial execution} is a finite prefix of some execution.  A (partial) execution
$ex'$ \emph{extends} some (partial) execution $ex$ if $ex$ is a prefix of $ex'$.
We say that a benign process~$p$ is \emph{correct} in an execution $ex$ if $p$ takes an infinite number of steps of $A$ in $ex$.  Otherwise a benign process $p$ is \emph{crash-faulty}.
We say that a \emph{crash-faulty} process $p$ \emph{crashes} at step $sp$ in an execution, if $sp$ is the last step of $p$ in that execution.

All writers in our model are benign and any number of them can be crash-faulty. Moreover, any number of readers can be Byzantine. Unless stated differently, we assume that up to $\td$ (resp., $\tm$) data (resp., metadata) replicas can be Byzantine; all other replicas are correct. Unless stated differently, we assume $D=2\td+1$ and $M=3\tm+1$.



\newcommand{\opread}{\textsc{read}}
\newcommand{\opwrite}{\textsc{write}}

\noindent\textbf{Storage object.} A storage abstraction is a shared \opread/\opwrite\ object.
Its sequential specification consists of a shared variable $x$ with two operations: \opwrite($v$), which takes a value
$v$ from domain $V$, stores $v$ in $x$ and returns special value $ok\notin V$, and \opread(), which returns the value of  $x$. We
assume that the initial value of $x$ is a special value
$\bot\notin V$.

We assume that each client invokes at most one operation at
a time (i.e., does not invoke the next operation until it receives
the response for the current operation). Only writers invoke \opwrite\
operations, whereas any client can invoke \opread\ operations. When we  talk about SWSR storage (single-writer single-reader), we assume that the writer and the reader are distinct process. Otherwise, we assume MWMR storage with $W \geq 1$ and $R \geq 1$.

For presentation simplicity, we do not explicitly model the initial
state of processes nor the invocations and responses of the
operations of the implemented storage object. We assume that
the algorithm $A$ initializes the processes in executions and determines
the invocations and responses of operations.
We say that $p$ \emph{invokes} an operation~$op$ at step $sp$ when $A$
modifies the state of a process $p$ in step $sp$ to start~$op$;
similarly, $op$ \emph{completes} at the step of $A$ when the response
of $op$ is received.


We say that a \opread/\opwrite\ operation $op$ is \emph{complete} in a (partial) execution if
the execution contains a response step for $op$. In any run, we say that a
complete operation $op_1$ \emph{precedes} operation $op_2$ (or $op_2$
\emph{follows} $op_1$) if the response step of $op_1$ precedes the
invocation step of $op_2$ in that run. If neither $op_1$ nor $op_2$
precedes the other, the operations are said to be \emph{concurrent}.

\newcommand{\optsread}{\textsc{tsread}}
\newcommand{\optswrite}{\textsc{tswrite}}

\noindent\textbf{Timestamped storage.} We use a special storage variant called \emph{timestamped storage} with
a slightly different sequential specification. Besides $x$ (initialized to $\bot$), timestamped storage maintains a timestamp $TS$ (an integer, initially $0$).
Timestamped storage exports the following operations:
\begin{itemize}
\item \optswrite(($ts,v$)) takes a pair of an integer timestamp~$ts$ and a value $v\in V$; \underline{if $ts\ge TS$}, then it stores $ts$ to $TS$ and $v$ to $x$ atomically\footnote{Here, in the sequential specification of \emph{timestamped storage}, it is critical to notice that the guard for a \optswrite\ to ``take effect'' requires $ts$ to be \emph{greater or equal} to $TS$. With such a condition, \emph{timestamped storage} has consensus number \cite{Her91} \emph{one}, and can be implemented with SWMR atomic registers as we discuss in Section~\ref{ssec:mds}. In contrast, \cite{CachinJS12} defines a ``replica'' object that is exactly the same as \emph{timestamped storage} except that the guard for the conditional write requires $ts$ to be \emph{strictly greater} than $TS$; this object, however, has consensus number $\infty$.}.  Regardless of timestamp $ts$, \optswrite\ returns $ok$.
\item \optsread() returns the pair ($TS,x$).
\end{itemize}


\noindent\textbf{Safety and liveness.} An algorithm \emph{implements} \emph{safe} (or \emph{atomic}) storage if every (partial) execution of the algorithm satisfies \emph{safety} (or \emph{atomi\-city}, respectively)
properties \cite{Lam86}.
We define safe storage for a single writer only
and say that a partial execution satisfies \emph{safety} if every \opread\
operation~$rd$ that is not concurrent with any \opwrite\ operation
returns value $v$ written by the last \opwrite\ that precedes $rd$, or $\bot$ in case there is no such \opwrite.
An execution $ex$ satisfies atomicity (or linearizability \cite{HW90})
if $ex$ can be extended (by appending zero or more response events) to an execution $ex'$ and there is a sequential permutation $\pi$ of $ex'$ (without incomplete invocations) such that $\pi$ preserves the real-time precedence order of operations in $ex$ and satisfies the sequential specification.
Moreover, a storage algorithm is \emph{obstruction-free} or \emph{wait-free} if every execution satisfies \emph{obstruction-freedom}~\cite{Herlihy:2003:OSD} or \emph{wait-freedom}~\cite{Her91}, respectively. Obstruction-freedom states that if a correct client invokes operation $op$ and no other client takes steps, $op$ eventually completes. Wait-freedom states that if a correct client invokes operation $op$, then $op$ eventually completes.
Atomicity and wait-freedom also apply to timestamped storage.

\newcommand{\MDSD}{MDS_{dir}}
\newcommand{\MDSH}{MDS_{hash}}
\newcommand{\mssg}[1]{\textsf{#1}}

\section{Protocol \protocol}
\label{sec:protocol}

In this section, we first give an overview of \protocol\ and then explain its modular pseudocode.
We then discuss possible implementations of the \protocol\ metadata service module using existing BFT storage protocols.
For lack of space, a full correctness proof is postponed to Appendix~\ref{app:proof}.

\subsection{Overview}
\label{ssec:overview}

\begin{algo*}[t]
\small
\newcounter{alg:client1:lines}
\centering
\begin{distribalgo}[1] \setcounter{ALC@line}{\value{alg:client1:lines}}
\smallskip

\INDENT {\textbf{Types:}}
\STATE \textit{TS}: $(\N_0 \times \{\N_0 \cup \bot\})$ with fields \textit{num} and \textit{cid} \hfil // timestamps
\STATE \textit{TSVals}: $(\textit{TS} \times \{\textit{V}$ $\cup \bot\})$ with fields \textit{ts} and \textit{val}
\STATE \textit{TSMeta}: $(\textit{TS} \times 2^{\N}) \cup \{\bot\}$ with fields \textit{ts} and \textit{replicas}
\ENDINDENT

\INDENT {\textbf{Shared objects:}}
\STATE $\MDSD$, is a \textit{MWMR} atomic wait-free timestamped storage object storing $x \in \textit{TSMeta}$
\STATE $\MDSH[ts\in TS]$ is an array of \textit{SWMR} safe wait-free storage objects storing $x \in H(V)$
\ENDINDENT

\INDENT {\textbf{Client state variables:}}
\STATE $md$: \textit{TSMeta}, initially $\bot$
\STATE $ts$: \textit{TS}, initially $(0,\bot)$
\STATE $Q :  2^{\N}$, initially $\emptyset$
\STATE $\textit{readval}$: $\textit{TSVals} \cup \{ \bot \}$, initially $\bot$
\ENDINDENT

\setcounter{alg:client1:lines}{\value{ALC@line}}
\end{distribalgo}
\begin{tabular}{c}\hline\mbox{}\hspace{0.97\textwidth}\mbox{}\end{tabular}
\begin{minipage}[t]{0.5\textwidth}
\begin{distribalgo}[1]  \setcounter{ALC@line}{\value{alg:client1:lines}}
\vspace{-1.5 em}
\INDENT {\textbf{operation} \textsc{write}$(v)$}
\STATE $md \leftarrow \MDSD.\optsread()$ \label{algwriter:tssyncbeg} \label{algwriter:mdsdread}
\IF {$md = \bot$}
\STATE $ts \leftarrow (0,c)$
\ELSE
\STATE $ts \leftarrow md.ts$  \label{algwriter:tssyncend}
\ENDIF
\smallskip
\STATE $ts \leftarrow (ts.num+1, c)$ \label{algclient:tsofwrite} \label{algwriter:writehashbeg}
\STATE $\MDSH[ts].\opwrite(H(v))$ \label{algwriter:mdshwrite} \label{algwriter:writehashend}
\smallskip
\STATE $Q \leftarrow \emptyset$  \label{algwriter:writedatabeg}
\INDENT {\textbf{for} $1 \leq i \leq D$ \textbf{do}} \label{algwriter:send2allbeg}
\STATE send \mssg{write}$\langle ts, v \rangle$ to $d_i$ \label{algwriter:send2allend}
\ENDINDENT
\STATE \textbf{wait until} $|Q| \geq t+1$ \label{algclient:wrcond}  \label{algwriter:writedataend}
\smallskip
\STATE $md \leftarrow (ts, Q)$ \label{algwriter:setmd} \label{algwriter:writedirbeg}
\STATE $\MDSD.\optswrite(md)$ \label{algwriter:mdsdwrite} \label{algwriter:writedirend}
\smallskip
\INDENT {\textbf{for} $1 \leq i \leq D$ \textbf{do}} \label{algwriter:commitbeg}
\STATE send \mssg{commit}$\langle ts \rangle$ to $d_i$ \label{algwriter:commitend}
\ENDINDENT
\smallskip
\STATE \textbf{return} \textsc{ok}
\ENDINDENT

\medskip
\INDENT {\textbf{upon} receiving \mssg{writeAck}$\langle ts \rangle$ from replica $d_i$}
\STATE $Q \leftarrow Q \cup \{i\}$
\ENDINDENT

\setcounter{alg:client1:lines}{\value{ALC@line}}
\end{distribalgo}
\end{minipage}%
\hfill
\begin{minipage}[t]{0.5\textwidth}
\begin{distribalgo}[1]   \setcounter{ALC@line}{\value{alg:client1:lines}}
\vspace{-1.5 em}

\INDENT {\textbf{operation} \textsc{read}$()$}
\STATE $\textit{readval} \leftarrow \bot$ \label{algreader:mdrdbeg}
\STATE $md \leftarrow \MDSD.\optsread()$ \label{algreader:mdsdread} \label{algreader:mdrdend}
\smallskip
\IF {$md = \bot$}
\STATE return $\bot$
\ENDIF

\INDENT {\textbf{for} $i \in md.replicas$ \textbf{do}}   \label{algreader:rdbeg}
\STATE send \mssg{read}$\langle md.ts \rangle$ to $d_i$ \label{algreader:rdsend}
\ENDINDENT
\STATE \textbf{wait until} $\textit{readval} \neq \bot$ \label{algclient:readcond}
\smallskip
\STATE \textbf{return} $\textit{readval}.val$ \label{algreader:rdend}
\ENDINDENT
\smallskip
\INDENT {\textbf{upon} receiving \mssg{readVal}$\langle ts', v' \rangle$ from replica $d_i$} \label{algclient:uponclause} \label{algreader:valbeg}
\IF {$\textit{readval} = \bot$}
    \IF  {$ts' = md.ts$} \label{algclient:uponcondeq} \label{algclient:valtsmatch}
        \STATE \textsc{check($ts',v'$)}
    \ENDIF
    \IF {$ts' > md.ts$} \label{algclient:valtshigh}
        \STATE $md' \leftarrow \MDSD.\optsread()$ \label{algclient:valtscheck} \label{algreader:mdsdread2}
        \IF {$md'.ts \geq ts'$}
            \STATE \textsc{check($ts',v'$)}
        \ENDIF
    \ENDIF
\ENDIF
\ENDINDENT

\smallskip
\INDENT {\textbf{procedure} \textsc{check}($ts,v$)} \label{algreader:checkbeg}
\IF {$H(v) = \MDSH[ts].\opread()$} \label{algreader:checkhash}
\STATE $\textit{readval} \leftarrow (ts,v)$ \label{algclient:tsofread} \label{algreader:valend} \label{algreader:checkend}
\ENDIF
\ENDINDENT

\setcounter{alg:client1:lines}{\value{ALC@line}}
\end{distribalgo}

\end{minipage}%

\caption{{Algorithm of client $c$.}}\label{alg:client}
\end{algo*}


\begin{algo*}[ht]
\small
\centering
\begin{tabular}{c}\hline\mbox{}\hspace{0.97\textwidth}\mbox{}\end{tabular}
\begin{minipage}[ht]{0.5\textwidth}
\begin{distribalgo}[1] \setcounter{ALC@line}{\value{alg:client1:lines}}
\vspace{-1 em}
\INDENT {\textbf{Server state variables:}}
\STATE $data$: $2^{\textit{TSVals}}$, initially $\emptyset$
\STATE $ts$: $\textit{TS}$, initially $(0, \bot)$
\ENDINDENT
\setcounter{alg:client1:lines}{\value{ALC@line}}
\end{distribalgo}
\noindent\rule{\columnwidth}{0.4pt}
\begin{distribalgo}[1]  \setcounter{ALC@line}{\value{alg:client1:lines}}
\vspace{-1 em}
\INDENT {\textbf{upon} receiving \mssg{read}$\langle ts' \rangle$ from client $c$} \label{replica:readcond} \label{algreplica:rdbeg}
\STATE \textbf{if} $ts' < ts$ \textbf{then} $ts' \leftarrow ts$ \label{algreplica:atleastts}
\STATE $v \leftarrow v' \in \textit{V} : (ts', v') \in data$
\STATE send \mssg{readVal}$\langle ts', v \rangle$ to $c$ \label{algreplica:rdend}
\ENDINDENT
\setcounter{alg:client1:lines}{\value{ALC@line}}
\end{distribalgo}
\end{minipage}%
\hfill
\begin{minipage}[ht]{0.5\textwidth}
\begin{distribalgo}[1]   \setcounter{ALC@line}{\value{alg:client1:lines}}
\vspace{-1 em}
\INDENT {\textbf{upon} receiving \mssg{write}$\langle ts', v' \rangle$ from client $c$} \label{algreplica:writebeg}
\IF {$ts' > ts$}
\STATE $data \leftarrow data \cup \{ (ts', v') \}$  \label{algreplica:writeadd}
\ENDIF
\STATE send \mssg{writeAck}$\langle ts' \rangle$ to client $c$ \label{algreplica:writeend}
\ENDINDENT

\smallskip

\INDENT {\textbf{upon} receiving \mssg{commit}$\langle ts' \rangle$ from client $c$}  \label{algreplica:commitbeg}
\IF {$ts' > ts \wedge \exists (ts', \cdot) \in data$}
\STATE $ts \leftarrow ts'$ \label{algreplica:setts}
\STATE $data \leftarrow data \setminus \{ (ts', \cdot) \in data : ts' < ts\}$  \label{algreplica:commitend}
\ENDIF
\ENDINDENT

\setcounter{alg:client1:lines}{\value{ALC@line}}
\end{distribalgo}

\end{minipage}%

\caption{{Algorithm of data replica $d_i$.}}\label{alg:replica}
\end{algo*}

\noindent
\protocol\ emulates multi-writer multi-reader (MWMR) atomic wait-free BFT storage, using $2t+1$ data replicas and $3t_M+1$ metadata replicas. Our implementation of \protocol\ is modular. Namely, metadata replicas are hidden within a \emph{metadata service} module $MDS$ which consists of: (a) a MWMR atomic wait-free timestamped storage object (denoted by $\MDSD$), which stores the metadata about the latest authoritative storage timestamp $ts$ and acts as a \emph{directory} by pointing to a set of $t+1$ data replicas that store the value associated with the latest timestamp (in the vein of  \cite{Farsite,FL03}); and (b) an array of SWMR safe wait-free storage objects (denoted by $\MDSH$), which each stores a hash of a value associated with a given timestamp $ts$, i.e., timestamps are used as indices for the $\MDSH$ array.  Every client may write to and read from $\MDSD$, but the entries of $\MDSH$ are written only once by a single client. Timestamps in \protocol are classical multi-writer timestamps~\cite{AW98,CachinGR11}, comprised of an integer $num$ and a process identifier $cid$ that serves to break ties.  Their comparison uses lexicographic ordering such that $ts_1>ts_2$ if and only if $ts_1.num>ts_2.num$ or $ts_1.num=ts_2.num$ and $ts_1.cid>ts_2.cid$.

The \protocol\ client pseudocode is given in Algorithm~\ref{alg:client} with data replica pseudocode given in Algorithm~\ref{alg:replica}. On a high level, \opwrite($v$) proceeds as follows: (i) a writer $w$ reads from $\MDSD$ to determine the latest timestamp $ts$ (Alg.~\ref{alg:client}, lines~\ref{algwriter:tssyncbeg}--\ref{algwriter:tssyncend}); (ii) $w$ increments $ts$ and writes the hash of value $v$ to $\MDSH[ts]$ (Alg.~\ref{alg:client}, lines~\ref{algwriter:writehashbeg}--\ref{algwriter:writehashend}); (iii) $w$ sends a \mssg{write} message to all data replicas containing $(ts,v)$ and waits for a set $Q$ of $t+1$ data replicas to reply (Alg.~\ref{alg:client}, lines~\ref{algwriter:writedatabeg}--\ref{algwriter:writedataend});
  (iv) $w$ writes $(ts,Q)$ to $\MDSD$ where $Q$ is a set of $t+1$ data replicas that have responded previously (Alg.~\ref{alg:client}, line~\ref{algwriter:writedirbeg}--\ref{algwriter:writedirend});
   and (v) $w$ sends a \mssg{commit} message to allow data replicas to garbage collect the data with timestamp less than $ts$ (Alg.~\ref{alg:client}, lines~\ref{algwriter:commitbeg}--\ref{algwriter:commitend}).

On the other hand, a reader $r$ upon invoking a \opread: (i) reads from $\MDSD$ the latest authoritative metadata $md$ with latest timestamp $md.ts$ and a set $md.replicas$ containing the identifiers of $t+1$ data replicas that store the latest value (Alg.~\ref{alg:client}, lines~\ref{algreader:mdrdbeg}--\ref{algreader:mdrdend}); and (ii) sends a \mssg{read} message to $md.replicas$ to read timestamp/value pairs not older than $md.ts$.
 Since clients do not trust replicas, reader $r$ needs to \emph{validate} every timestamp/value received in a \mssg{readVal} message sent by a data replica in response to a \mssg{read} message (Alg.~\ref{alg:replica}, lines~\ref{algreplica:rdbeg}--\ref{algreplica:rdend}).
 To this end, readers consult the metadata service (Alg.~\ref{alg:client}, lines~\ref{algreader:valbeg}--\ref{algreader:valend}): (i) in case the timestamp received from a data replica $ts'$ equals the timestamp in $md.ts$ (Alg.~\ref{alg:client}, line~\ref{algclient:valtsmatch}) then the reader only checks whether the value has indeed been written by reading $\MDSH[md.ts]$ and comparing this to the hash of the received value; otherwise (ii), i.e., when $ts'>md.ts$ (Alg.~\ref{alg:client}, line~\ref{algclient:valtshigh}), the reader first validates $ts'$ itself by checking if $\MDSD$ points to $ts'$ or even a later timestamp, and, if yes, proceeds to check the integrity of the value by comparing its hash to the value in~$\MDSH[ts']$.


\subsection{\protocol\ details}
\label{ssec:pseudocode}

\begin{figure}[bh]
\begin{center}
        \includegraphics*[width=\textwidth]{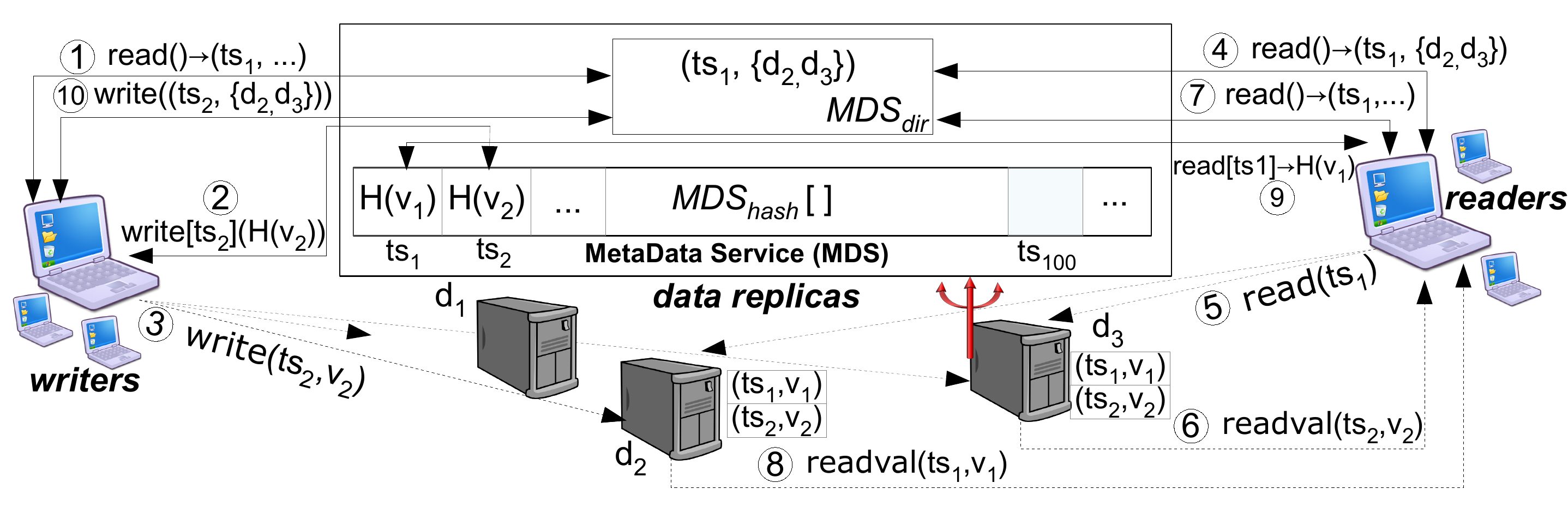}
\end{center}
\vspace{-2 em}
 \caption{Illustration of a \protocol\ execution with a concurrent \opwrite\ and \opread.}
 \label{fig:run}
\end{figure}

We further illustrate \protocol\ using an execution $ex$, depicted in Figure~\ref{fig:run}. In $ex$, we assume $t=1$ and hence $D=3$ data replicas. In $ex$, data replica $d_1$ due to asynchrony does not receive messages in a timely manner, whereas data replica $d_3$ is Byzantine.

Execution $ex$ starts with a complete $wr_1=\opwrite(v_1)$ which stores $(ts_1,v_1)$ into data replicas $\{d_2,d_3\}$, where $ts_1$ is a pair $(1,w_1)$ that writer $w_1$ generated in line~\ref{algclient:tsofwrite} of $wr_1$. Notice that \opwrite\ $wr_1$ is not explicitly shown in Figure~\ref{fig:run}; however, the states of $\MDSD$ and $\MDSH[ts_1]$ upon completion of $wr_1$ are shown.

In $ex$, the initial $wr_1$ is followed by two concurrent operations depicted in Figure~\ref{fig:run}: (i) a $wr_2=\opwrite(v_2)$ by writer $w_2$, and (ii) \opread\ $rd$ by reader $r_1$. Upon invoking $wr_2$, writer $w_2$ in Step \circled{1} (we refer to numbers in Fig.~\ref{fig:run}) first reads from $\MDSD$ the latest timestamp by invoking $\MDSD.\optsread()$ (line~\ref{algwriter:tssyncbeg}). $\MDSD$ eventually responds and $w_2$ reads timestamp $ts_1=(1,w_1)$. Then, writer $w_2$ increments the timestamp and adds its own identifier (line~\ref{algclient:tsofwrite}) to obtain timestamp $ts_2=(2,w_2)$.  Then, writer $w_2$ invokes $\MDSH[ts_2].\opwrite(H(v_2))$  where $H(v_2)$ is a hash of written value $v_2$ (line~\ref{algwriter:mdshwrite}, Step \circled{2}). Values written to $\MDSH$ serve to ensure integrity in the presence of potentially Byzantine data replicas; a writer writes to $\MDSH$ before exposing the current \opwrite\ to other clients  by writing to $\MDSD$ in order to prevent Byzantine replicas forging values with a given timestamp. Eventually, $\MDSH$ responds and writer $w_2$ then sends a \mssg{write}$\langle ts_2, v_2 \rangle$ message to all data replicas containing an entire value $v_2$ (lines~\ref{algwriter:send2allbeg}--\ref{algwriter:send2allend}, Step \circled{3}). In $ex$, \mssg{write} messages are received only by data replicas $d_2$ and $d_3$ (which is Byzantine). A correct replica $d_2$ simply adds the pair ($ts_2,v_2$) to its $data$ set (line~\ref{algreplica:writeadd}) but $d_2$ does not update its local authoritative timestamp $ts$ which still reflects $ts_1$. At this point in time of execution $ex$, we make writer $w_2$ wait for asynchronous replies from data replicas.

At the same time, concurrently with $wr_2$, reader $r_1$ invokes \opread\ $rd$. Reader $r_1$ first queries $\MDSD$ for metadata $md$ by invoking $\MDSD.\optsread()$, to determine the latest timestamp $md.ts$ and the set of data replicas $md.replicas$ that store the latest value (line~\ref{algreader:mdsdread}, Step \circled{4}). $\MDSD$ eventually responds and $r_1$ sees $md.ts=ts_1$ and $md.replicas=\{d_2,d_3\}$. Then, $r_1$ sends \mssg{read} message to data replicas $d_2$ and $d_3$ (lines~\ref{algreader:rdbeg}--\ref{algreader:rdsend}, Step \circled{5}).
By the algorithm, a data replica~$d$ replies to a \mssg{read} message with a \mssg{readVal} message containing the value associated with its local authoritative timestamp~$ts$, which does not necessarily reflect the highest timestamp that replica~$d$ stores in $data$; e.g., in case of $d_2$ (and $d_3$) in $ex$, $ts$ equals $ts_1$ and not $ts_2$. 
However, a Byzantine $d_3$ could mount a sophisticated attack and respond with the pair $(ts_2,v_2)$ (Step \circled{6}); although this pair is in fact written concurrently, it is dangerous for $r_1$ to return $v_2$ since, in \protocol\, readers do not write back data and the value $v_2$ has not been completely written~--- this may violate atomicity. To prevent this attack, a reader invokes $\MDSD.\optsread()$ to determine whether $ts_2$ (or a higher timestamp) became authoritative in the mean time (lines~\ref{algclient:valtshigh}--\ref{algclient:valtsmatch}, Step \circled{7}). Since this is not the case, $r_1$ discards the reply from $d_3$ and waits for an additional reply (from $d_2$).

An alternative attack by Byzantine $d_3$ could be to make up a timestamp/value pair with a large timestamp, say $ts_{100}$. In this case, $r_1$ would also first check with $\MDSD$  whether $ts_{100}$ or a higher timestamp has been written (just like in Step \circled{7}). However, if so, $r_1$ would then proceed to check the integrity of a hash of the value reported by $d_3$ by invoking $\MDSH[ts_{100}].\opread()$ (lines~\ref{algreader:checkbeg}--\ref{algreader:checkend}); this check would assuming a bounded adversary as the hash function is collision-free.

In $ex$, eventually $d_2$ responds to $r_1$ with pair $(ts_1,v_1)$ (lines~\ref{algreplica:rdbeg}--\ref{algreplica:rdend}, Step \circled{8}). By the protocol (optimizations omitted for clarity) reader $r_1$ verifies the integrity of $v_1$ by reading a hash from $\MDSH[ts_{1}]$ (lines~\ref{algreader:checkbeg}--\ref{algreader:checkend}, Step \circled{9}). This time, the check succeeds and $rd$ completes returning value $v_1$.

Eventually, writer $w_2$ receives \mssg{writeAck} replies in $wr_2$ from replicas $d_2$ and $d_3$. Then, writer $w_2$  invokes $\MDSD.\optswrite(ts_2,\{d_2,d_3\})$ (lines~\ref{algwriter:writedirbeg}--\ref{algwriter:writedirend}, Step \circled{10}) only now, when the write~$wr_2$
 finally ``takes effect'', i.e., at the linearization point of \opwrite\ which coincides with the linearization point of the \optswrite\ to $\MDSD$. Finally, the writer sends a \mssg{commit} message to all replicas to allow them to garbage collect stale data (lines~\ref{algwriter:commitbeg}--\ref{algwriter:commitend}); notice that data replicas update their local variable $ts$, which reflects a value they will serve to a reader, only upon receiving a \mssg{commit} message (lines~\ref{algreplica:commitbeg}--\ref{algreplica:commitend}).

Finally, we point out that \protocol\ uses timestamped storage ($MDS_{dir}$) as a way to avoid storing entire history of a shared variable at data replicas. We could not achieve this with $MDS_{dir}$ being a classical storage object, since such a classical storage object would allow overwrites of $MDS_{dir}$ with a lower timestamp. With our protocol at data replicas (notably lines~\ref{algreplica:writebeg}--\ref{algreplica:writeend}) and our goal of not storing entire histories, such an overwrite could put $MDS_{dir}$ in inconsistent state with data replicas.

\subsection{Metadata service implementations}
\label{ssec:mds}

We show how to implement the \protocol\ metadata service from existing asynchronous BFT storage protocols that rely on $3t+1$ replicas~--- in our case these are exactly $3t_M+1$ metadata replicas. To qualify for reuse, existing BFT protocols should also tolerate an arbitrary number of Byzantine readers, any number of crash-faulty writers, and, ideally, make no  cryptographic assumptions.

First, it is critical to see that $\MDSD$, our MWMR atomic wait-free timestamped storage, can be implemented as a straightforward extension of the classical SWMR to MWMR atomic storage object transformation (e.g., \cite[page~163]{CachinGR11}). In this transformation, there is one SWMR storage object per writer and writers store timestamp/value pairs in ``their'' storage object, after first reading and incrementing the highest timestamp found in any other storage object. In this extension, the reader determines the timestamp/value pair with the highest timestamp among the SWMR storage objects as usual, and simply returns also the timestamp together with the value.
%
This implementation may be realized from existing SWMR atomic wait-free storage (with $3t+1$ replicas); examples include \cite{AAB07,PoW} (with an unbounded adversary) and \cite{Phalanx,Cachin:2006:ORE} (with a bounded adversary).

Second, $\MDSH$ is an array of SWMR safe storage objects that can directly be implemented from the protocols with atomic semantics mentioned above, or even from protocols with weaker implementations, such as (i) SWMR safe wait-free storage \cite{ACKM06} or (ii) its regular variant, both without cryptographic assumptions~\cite{GV06}, or (iii)  regular storage with digital signatures~\cite{MR98}.

Finally, we add that more efficient, direct, implementations of the \protocol\ metadata service can be obtained easily, but these are beyond the scope of this paper.


\section{Lower bounds}
\label{sec:lb}

\newcommand{\available}{obstruction-free}
\newcommand{\availability}{obstruction-freedom}
\newcommand{\consistent}{safe}
\newcommand{\consistency}{safety}

In this section we prove two lower bounds: (i) we show that using $2\td+1$ data replicas to tolerate $\td$ data replica \emph{crash} faults is necessary implementing distributed single-writer single-reader \available\ \consistent\ storage, even with the help of a \emph{metadata service oracle}; and (ii) we also show that the same result extends to $3\td+1$ replicas in the model with Byzantine data replicas. However, this second lower bound applies in the model with an \emph{unbounded adversary} and does not hold when the adversary is \emph{bounded}, i.e., when it cannot break cryptographic hash functions (see Sec.~\ref{sec:model}).

Technically, we unify the two results into one single argument in a \emph{hybrid failure} model, where we consider $D=2\td+\bd+1$ data replicas, out of which up to $\bd$ can be Byzantine and up to $\td-\bd$ can only crash. For the purpose of this proof, we focus on the model with a single writer.\\

\label{def:lbprel}
\noindent\textbf{Preliminaries.} Our lower bound model assumes a \emph{metadata service} (Def.~\ref{def:mds}): in short, a metadata service is an oracle, modeled as a \emph{correct} process.\footnote{In our proof we do not use metadata replicas (defined in Section~\ref{sec:model}) which are ``replaced'' by a metadata service oracle.} A metadata service is parameterized by the domain of values $V$ of the implemented storage. Roughly speaking, a metadata service can implement an arbitrary functionality, except that it might not be able to help a reader distinguish whether the writer wrote value $v\in V$ or value $v'\in V$, where $v\neq v'$.

\begin{definition} [Metadata service]
\label{def:mds}
A metadata service for a value domain $V$ (denoted by $\textit{MDS}_V$) is a \emph{correct} process that can implement an arbitrary automaton with the following limitation.  There exist two values $v, v'\in V$ (we say they are \emph{indistinguishable to $\textit{MDS}_V$}), such that there is no distributed storage algorithm for $\textit{MDS}_V$, the writer and a set of processes $P$, such that some process $p\in P$ can distinguish execution $ex_v$ from $ex_{v'}$, where:
\begin{itemize}
\item In $ex_v$, the writer invokes a complete $\textsc{write}(v)$ and crashes, such that no process in $P$ receives any message from the writer in $ex_v$; and
\item In $ex_{v'}$, the writer invokes a complete $\textsc{write}(v')$ and crashes, such that no process in $P$ receives any message from the writer in $ex_{v'}$.
\end{itemize}
\end{definition}

Intuitively, Definition~\ref{def:mds} models metadata service as a general oracle with arbitrary functionality, with the restriction that it cannot store or relay data values in $V$. Observe that if we extend both executions $ex_v$ and $ex_{v'}$ in Definition~\ref{def:mds} by appending a \opread\ by a correct reader (from $P$), to obtain partial executions $ex'_v$ and $ex'_{v'}$, respectively, \availability\ or \consistency\ is violated in $ex'_v$ or in $ex'_{v'}$.

To state our lower bound precisely, we change the model of Section~\ref{sec:model} and assume that, out of $D$ data replicas, up to $\bd$ can be Byzantine and  additionally $\td-\bd$ of them are benign (that is, they may crash), for $0\le \bd\le \td$.
We assume an \emph{unbounded adversary} that can coordinate Byzantine processes and that either knows values $v$ and $v'$ that are indistinguishable to $\textit{MDS}_V$, or can compute such a $v$ given $v'$, or vice-versa.

We now state the main result of this section:

\begin{theorem}
\label{the:crash}
Assuming an umbounded adversary, there is no asynchronous distributed
algorithm that implements single-writer single-reader (SWSR)
\available\ \consistent\ storage (with domain $V$), with $D\le
2\td+\bd$ data replicas and a metadata service for $V$.
\end{theorem}


\begin{prooff}
Assume by contradiction that such implementation~$I$ exists. We develop a series of executions of $I$ 
to show that at most $2\td+\bd$ data replicas do not help the reader distinguish the values indistinguishable to $\textit{MDS}_V$, $v$ and $v'$. To this end, we divide the set of data replicas in three disjoint \emph{groups} $T_1$ and $T_2$, each containing at most $\td$ data replicas, and group $B$ with at most $\bd$ data replicas.

Consider first partial execution $ex_1$ in which the reader and the replicas from group $T_1$ crash at the beginning of $ex_1$ and the writer invokes $\opwrite(v)$. By \availability\, $wr_1$ eventually completes. Then, the writer crashes and $ex_1$ ends at time $t_1$. In $ex_1$, the reader and data replicas from $T_1$ do not deliver any message, whereas the writer, $\textit{MDS}_V$ and data replicas from $T_2\cup B$ deliver all the messages per implementation $I$. We denote the state of data replicas from group $B$ at $t_1$ by $\sigma_v$.

Let $ex_2$ be a partial execution in which the writer invokes $\textsc{write}(v')$ that ends at time $t_2$, such that $ex_2$ is otherwise similar to $ex_1$, with the reader and replicas from $T_1$ crashing at the beginning of $ex_2$ and the other processes delivering all messages. We denote the state of data replicas from group $B$ at $t_2$ by $\sigma_{v'}$.

Let $ex'_1$ be a partial execution similar to $ex_1$, except that the reader and the data replicas from $T_1$ do not crash, yet they still do not receive any message by time $t_1$ (due to asynchrony). At time $t_1$, data replicas from $T_2$ crash. This is followed by \opread\ $rd_1$ by the reader at $t_3>max\{t_1,t_2\}$. The reader and data replicas in $T_1$, never receive any message from faulty data replicas from $T_2$ or the faulty writer. By \availability\, $rd_1$ eventually completes (at $t_4$) and, by \consistency, returns the value written by $wr_1$, i.e., $v$.

Let $ex'_2$ be a partial execution similar to $ex_2$, except that the reader and the replicas from $T_1$ are not faulty, yet they still do not receive any message by time $t_2$ (due to asynchrony). At time $t_2$, data replicas from $B$ (if any) exhibit a Byzantine fault, by \emph{changing their state from $\sigma_{v'}$ to $\sigma_{v}$} (see $ex_1$). After this, data replicas from $B$ follow the protocol. This is followed by a \opread\ $rd_2$ by the reader at $t_3$. Moreover, assume that due to asynchrony, $\textit{MDS}_V$, the reader and data replicas in $T_1$, do not receive any message from data replicas from $T_2$ until after $t_4$. Notice that, by Definition~\ref{def:mds} and since they do not receive any message from the writer or data replicas in $T_2$, the reader and the data replicas in $T_1$ cannot distinguish $ex'_2$ from $ex'_1$. Hence, in $ex'_2$, $rd_2$ returns $v$ (at $t_4$) like in $ex'_1$. However, this violates \consistency\ by which $rd_2$ must return $v'$. A contradiction.
\end{prooff}

\noindent\textbf{Discussion.} We make two observations about Theorem~\ref{the:crash}. First, in the crash model, where $\bd=0$, Theorem~\ref{the:crash} implies that $2\td+1$ data replicas are necessary for implementing SWSR \available\ \consistent\ storage, even with a metadata service oracle. Second, notice that the Byzantine part of the proof critically relies on the ability of the adversary to successfully switch from the state where Byzantine replicas ``observed'' $v'$ to the state where Byzantine replicas seemingly have ``observed'' $v$ (see $ex'_2$). In practice, when assuming a bounded adversary, cryptographic hash functions easily prevent this attack~--- the proof of Theorem~\ref{the:crash} breaks down for $\bd>0$. Protocols in this realistic model, including \protocol, are only subject to the lower bound of $2\td+1$ data replicas from the crash model. 

\omitit{
\subsection{Discussion}

We make two observations about Theorem~\ref{the:crash} and its proof in Section~\ref{sec:proof}.
\begin{enumerate}
\item In the crash model, where $\bd=0$, Theorem~\ref{the:crash} implies that $2\td+1$ data replicas are necessary for implementing SWSR \available\ \consistent\ storage, even with a metadata service oracle.

\item Byzantine ``part'' of the proof critically relies on the ability of the adversary to successfully switch from the state where Byzantine replicas ``observed'' $v'$ to the state where Byzantine replicas seemingly have ``observed'' $v$ (see $ex'_2$, Sec.~\ref{sec:proof}). In practice this can be easily prevented using collision resistant cryptographic hash functions. In this case, our lower bound proof and Theorem~\ref{the:crash} do not apply for $\bd>0$. Implementations in models with such a bounded (yet very realistic) adversary, are only subject to the $2\td+1$ lower bound from the crash model. We demonstrated one such implementation with our \protocol\ protocol in Section~\ref{sec:protocol}.
\end{enumerate} }

\section{Related work}
\label{sec:relwork}

The read/write storage abstraction (also known as a \emph{register}) was formalized by Lamport \cite{Lam86}.  
Martin et al. \cite{MAD02} demonstrated a tight lower bound of $3t+1$ replicas needed for any register implementation that tolerates $t$ Byzantine replicas in an asynchronous system. Their bound applies even to single-writer single-reader safe register, where the reader and the writer are benign. In this paper, we refine this bound by logically separating storage replicas into data replicas and metadata replicas. With such a separation, we show that the $3t+1$ lower bound of \cite{MAD02} applies to register metadata replicas only, but it does not hold for the number of data  replicas. Only $2t+1$ data replicas are needed to tolerate $t$ Byzantine data replicas in an asynchronous system with a bounded adversary.

Protocol \protocol that matches this lower bound is similar in style to Farsite \cite{Farsite}, a BFT file service, and Hybris~\cite{Hybris}, a recent hybrid cloud storage system. Namely, like \protocol, Farsite and Hybris separate metadata from data and keep cryptographic hashes and the directory information as metadata and require at least $2t+1$ data replicas. However, unlike \protocol, Farsite and Hybris metadata services are based on replicated state machines; hence both Farsite and Hybris are subject to the FLP impossibility result [14] and require stronger timing assumptions, such as partial synchrony [11]. In addition, Farsite supports single-writer and multiple readers and uses read/write locks for concurrency control, whereas our \protocol\ supports multiple writers and offers wait-free \cite{Her91} atomic semantics, without resorting to locks. On the other hand, Hybris only supports FW-terminating reads and is not wait-free.

Data and metadata have also been separated in asynchronous \emph{crash-tolerant} storage~\cite{FL03,Vivace} and in variants of state-machine replication~\cite{Gnothi}.
Interestingly, separating data from metadata does not reap benefits in terms of reduced resource costs with crash-faults: indeed all of the mentioned crash-tolerant protocols
that exercise data/metadata separation \cite{FL03,Vivace,Gnothi} still need $2t+1$ data replicas. We prove this  inherent: even with a fault-free metadata service, $2t+1$ data replicas are necessary to tolerate $t$ data replica faults.

Separation of data from the control plane is well-known in consensus and state machine replication. Lamport's Paxos algorithm \cite{lampor98,PaxosMadeSimple} separated consensus roles into proposers, acceptors, and learners. In this context,
the lower bound of $3t+1$ replicas for partially synchronous BFT consensus was shown to apply only to acceptors \cite{Lam03}, but not to proposers or learners. For example, \cite{GV10} demonstrated a partially synchronous BFT consensus protocol in which any number of proposers and learners can be Byzantine.
Yin et al. \cite{Yin03} apply the ideas from Lamport's consensus role separation and separate \emph{agreement} from \emph{execution} to obtain state machine replication protocols 
with $3t+1$ agreement replicas and $2t+1$ execution replicas. However, just like \cite{Farsite,Hybris}, the results of
\cite{Yin03, GV10} that apply to state-machine replication and consensus are fundamentally different from ours; they are
subject to the FLP impossibility result \cite{FLP85} and the protocols
therefore rely on stronger timing assumptions~\cite{DLS88}.

\section{Conclusion and future work}
\label{sec:conclusion}

This paper presents \protocol, the first asynchronous BFT storage protocol that uses $2t+1$ data replicas to tolerate $t$ Byzantine faults in a general model without trusted components. To achieve this, \protocol
separates data from metadata  and stores metadata leveraging a novel abstraction we call \emph{timestamped storage} which  can can be implemented using existing asynchronous BFT storage protocols that need $3t+1$ replicas to tolerate $t$ Byzantine faults. In addition, \protocol implements strong guarantees such as wait-freedom and atomicity (linearizability). Finally, \protocol relies on collision-resistant cryptographic hash functions which we show inherent.
In this paper we show also that, perhaps surprisingly, no asynchronous crash-tolerant storage implementation can achieve better resilience with respect to data replicas than our BFT \protocol.

Our work opens many avenues for future work in BFT storage systems, especially for those of practical relevance.
It requires to revisit other important aspects of asynchronous BFT storage, such as their complexity or  erasure-coded implementations, which have been extensively studied in the traditional model with unified data and metadata.



\newpage
\appendix

\section{Correctness of \protocol}
\label{app:proof}
In this section we prove that the pseudocode in Algorithm~\ref{alg:client} and Algorithm~\ref{alg:replica} is correct by showing that it satisfies atomicity and wait-freedom.

\begin{definition} [Timestamp of an Operation] \label{def:tso}
If $o$ is an operation, then we define the timestamp of $o$, denoted $ts(o)$ as follows. If $o$ is a \textsc{write} operation,
then  $ts(o)$ is $ts$ when its assignment completes in line~\ref{algclient:tsofwrite}. Else, if $o$ is \textsc{read} operation,
then $ts(o)$ is the timestamp associated to $readval$ when its assignment completes in line~\ref{algclient:tsofread}.
\end{definition}

\begin{lemma}[Timestamped Storage Safety]~\label{la:safety}
Let $x$ be a timestamped storage object and let $tsrd$ be an operation $x.\optsread$ returning $(ts',v')$. If $tsrd$ follows after an operation $x.\optswrite((ts,v))$ or after an operation $x.\optsread$ returning $(ts,v)$, then $ts' \geq ts$.
\end{lemma}
\begin{prooff}
Follows from the sequential specification of timestamped storage.
\end{prooff}

\begin{lemma} [Sandwich] \label{la:sandwich}
 Let $rd$ be a complete \opread\ operation and let $md$ and $md'$ be the value returned by $\MDSD$ in lines~\ref{algreader:mdsdread}
 and~\ref{algreader:mdsdread2} respectively, Then $md.ts \leq ts(rd) \leq md'.ts$.
\end{lemma}
\begin{prooff}
By Definition~\ref{def:tso}, $ts(rd)$ is $\textit{readval}.ts$ when the assignment in line~\ref{algclient:tsofread} completes.
For this to happen, either the condition in line~\ref{algclient:valtsmatch} or line~\ref{algclient:valtshigh} must be satisfied.
This is implies that either $ts(rd) = md.ts$ or $md.ts < ts(rd) \leq md'.ts$.
\end{prooff}

\begin{lemma} [Partial Order] \label{la:po}
Let $o$ and $o'$ be two operations with timestamps $ts(o)$ and $ts(o')$, respectively, such that $o$ precedes $o'$.
Then $ts(o) \leq ts(o')$ and if $o'$ is a \textsc{write} operation then $ts(o) < ts(o')$.
\end{lemma}
\begin{prooff}
Let $o'$ be a \textsc{read} (resp. a \textsc{write}) operation. By Definition~\ref{def:tso} and Lemma~\ref{la:sandwich}, $ts(o') \geq o'.md.ts$ ($o.$ denotes the context of operation $o$). In the following we distinguish whether $o$ is a \opwrite\ or a \opread.

\noindent \textbf{Case 1} ($o$ is a \opwrite): if $o$ is a \opwrite, then $o.\MDSD.\optswrite(o.md)$ in line~\ref{algwriter:mdsdwrite}
precedes $o'.md \leftarrow o'.\MDSD.\optsread()$ in line~\ref{algreader:mdsdread} (resp. ~\ref{algwriter:mdsdread}).
By Lemma~\ref{la:safety}, it follows that $o'.md.ts \geq o.md.ts$. By Definition~\ref{def:tso} $o.md.ts = ts(o)$, and therefore $o'.md.ts \geq ts(o)$.
There are two possible subcases; either $o'$ is a \opread\ or a \opwrite. If $o'$ is a \opread\ then $ts(o') \geq o'.md.ts$, and therefore $ts(o') \geq ts(o)$.
Otherwise, if $o'$ is a \opwrite, then $ts(o') > o'.md.ts$ because $ts(o')$ is obtained from incrementing the first component of $o'.md.ts$.
Therefore, $ts(o') > o'.md.ts \geq ts(o)$.

\noindent \textbf{Case 2} ($o$ is a \opread): if $o$ is a \opread, then by Definition~\ref{def:tso} and Lemma~\ref{la:sandwich}, $o.md.ts \leq ts(o) \leq o.md'.ts$.
In what follows, we treat the only two possible cases $ts(o) = o.md.ts$ and $o.md.ts < ts(o) \leq o.md'.ts$ separately.

\indent \textbf{(2a)}: if $ts(o) = o.md.ts$, then since $o'.md \leftarrow o'.\MDSD.\optsread()$ in line~\ref{algreader:mdsdread}
(resp. ~\ref{algwriter:mdsdread}) follows after $o.md \leftarrow o.\MDSD.\optsread()$, by Lemma~\ref{la:safety} $o'.md.ts \geq o.md.ts$.
Since $o.md.ts = ts(o)$, it follows that $o'.md.ts \geq ts(o)$.
If $o'$ is a \opread, then $ts(o') \geq o'.md.ts$, and we conclude that $ts(o') \geq ts(o)$. Otherwise, if $o'$ is a \opwrite,
then $ts(o') > o'.md.ts$ and therefore, $ts(o') > ts(o)$.

\indent \textbf{(2b)}: if $o.md.ts < ts(o) \leq o.md'.ts$, then $o.md' \leftarrow o.\MDSD.\optsread()$ in line~\ref{algreader:mdsdread2}
precedes $o'.md \leftarrow o'.\MDSD.\optsread()$ in line~\ref{algreader:mdsdread} (resp. ~\ref{algwriter:mdsdread}).
By Lemma~\ref{la:safety} $o'.md.ts \geq o.md'.ts$ and since $o.md'.ts \geq ts(o)$, it follows that $o'.md.ts \geq ts(o)$.
If $o'$ is a \opread, then $ts(o') \geq o'.md.ts$, and we conclude that $ts(o') \geq ts(o)$.  Otherwise, if $o'$ is a \opwrite,
then $ts(o') > o'.md.ts$ and therefore, $ts(o') > ts(o)$, which completes the proof.
\end{prooff}

\begin{lemma}[Unique Writes]\label{la:writetimestamps}
  If $o$ and $o'$ are two \textsc{write} operations with timestamps $ts(o)$ and $ts(o')$, then $ts(o) \neq ts(o')$.
\end{lemma}
\begin{prooff}
  If $o$ and $o'$ are executed by different clients, then the two timestamps differ in their second component. If $o$ and $o'$ are executed by the same client, then
  the client executed them sequentially. By Lemma~\ref{la:po}, $ts(o') \neq ts(o)$.
\end{prooff}

\begin{lemma}[Integrity]\label{la:readintegrity}
  Let $rd$ be a \textsc{read} operation with timestamp $ts(rd)$ returning value $v \neq \bot$. Then there is a single \opwrite\ operation $wr$ of the form \textsc{write}($v$) such that $ts(rd) = ts(wr)$. 

\end{lemma}
\begin{prooff}
    Since $rd$ returns $v$ and has an associated timestamp $ts(rd)$,
    $rd$ receives $(ts(rd), v)$ from one of the data replicas.
    Suppose for the purpose of contradiction that $v$ is never written.
    Then, then by the collision resistance of $H$, the check in
    line~\ref{algreader:checkhash} does not pass and $rd$ does not return $v$.
    Therefore, we conclude that some operation $wr$ sends a message \textsf{write}$\langle ts(rd), v \rangle$
    in line~\ref{algwriter:send2allend}. Since $ts(wr)$ is set only once during the execution of a \opwrite\ and
    that occurs in line~\ref{algclient:tsofwrite}, it follows that $ts(wr) = ts(rd)$. Finally,
    by Lemma~\ref{la:writetimestamps} no other write has the same timestamp, which completes the proof.
\end{prooff}

\begin{theorem}[Atomicity (Linearizability)]
  Every execution $ex$ of Algorithm~\ref{alg:client} and Algorithm~\ref{alg:replica} satisfies atomicity.
\end{theorem}
\begin{prooff}
    Let $ex$ be an execution of the algorithm.  By Lemma~\ref{la:readintegrity}
    the timestamp of a \opread\ operation either has been written by
    some \opwrite\ operation or the \opread\ returns $\bot$.

    We first construct $ex'$ from $ex$ by completing all \textsc{write}
    operations of the form \textsc{write}$(v)$, where $v$ has been
    returned by some complete \textsc{read} operation. Then we construct
    a sequential permutation $\pi$ by ordering all operations in $ex'$ excluding the \opread\
    operations that did return $\bot$ according to their timestamps
    and by placing all \opread\ operations that did not return $\bot$ immediately
    after the \textsc{write} operation with the same timestamp. The \opread\ operations that
    did return $\bot$ are placed in the beginning of $\pi$. Note that concurrent
    \textsc{read} operations  with the same timestamp may appear in any order, whereas all
    other \textsc{read} operations appear in the same order as in $ex'$.

  To prove that $\pi$ preserves the sequential specification of a MWMR
  register we must show that a \textsc{read} always returns the value
  written by the latest preceding write which appears before it in $\pi$,
  or the initial value of the register $\bot$ if there is no preceding
  write in $\pi$.  Let $rd$ be a \textsc{read} operation returning a
  value $v$. If $v = \bot$, then by construction $rd$ is ordered before
  any \opwrite\ in $\pi$.

  Otherwise, $v \neq \bot$ and by Lemma~\ref{la:readintegrity}
  there exists a \opwrite$(v)$ operation, with the same
  timestamp, $ts(rd)$. In this case, this write is placed in
  $\pi$ before $rd$, by construction. By Lemma~\ref{la:writetimestamps},
  other write operations in $\pi$ have a different associated timestamp
  and therefore appear in $\pi$ either before \opwrite$(v)$ or after $rd$.

  It remains to show that $\pi$ preserves real-time order. Consider
  two complete operations $o$ and $o'$ in $ex'$ such that $o$ precedes
  $o'$. By Lemma~\ref{la:po}, $ts(o') \geq ts(o)$. If $ts(o') > ts(o)$
  then $o'$ appears after $o$ in $\pi$ by construction. Otherwise
  $ts(o') = ts(o)$ and by Lemma~\ref{la:po} it follows that $o'$ is a
  \opread\ operation. If $o$ is a \opwrite\ operation, then $o'$
  appears after $o$ since we placed each read after the \opwrite\
  with the same timestamp. Otherwise, if $o$ is a \opread, then
  it appears before $o'$ as in $ex'$.
\end{prooff}

\begin{theorem}[Wait-Freedom]
The protocol comprising Algorithm~\ref{alg:client} and Algorithm~\ref{alg:replica} satisfies wait-freedom.
\end{theorem}
\begin{prooff}
Since the shared storage objects used in Algorithm~\ref{alg:client} are wait-free, every \opread\ or \opwrite\ operation invoked on $\MDSD$ and $\MDSH[ts]$, where $ts \in TSVals$, eventually completes. It remains to show that no \textsc{write} (resp. \textsc{read}) operation blocks in line~\ref{algclient:wrcond} (resp.~\ref{algclient:readcond}).  For a \opwrite\ operation $wr$, the waiting condition in line~\ref{algclient:wrcond} is eventually satisfied because there is a time after which all correct data replicas reply and there are at least $t+1$ such replicas. On the other hand, let $rd$ be a $\textsc{read}$ operation and suppose for the purpose of contradiction that the waiting condition in line~\ref{algclient:readcond} is never satisfied, and therefore \textit{readval} is never set in line~\ref{algclient:tsofread}. Let $d_i$ be a correct data replica such that $i \in md.replicas$. Since $rd$ did previously sent a $\mssg{read}\langle md.ts \rangle$ message to $d_i$, eventually $rd$ receives a reply from $d_i$ consisting of a pair $(ts',v)$ in line~\ref{algclient:uponclause}.

If $ts'$ satisfies $md.ts \leq ts' \leq md'.ts$, then since $d_i$ is a correct replica, the condition in line~\ref{algreader:checkhash} is also satisfied, and therefore \textit{readval} is set in line~\ref{algclient:tsofread}. Suppose for the purpose of contradiction that $ts' < md.ts$ or $ts' > md'.ts$.
Notice that the requested timestamp is $md.ts$. If $ts' < md.ts$ then $d_i$ replied with a smaller timestamp than $md.ts$. However, notice that according to the check in the replica code in line~\ref{algreplica:atleastts}, $d_i$ never replies with a timestamp smaller than the requested timestamp, contradicting our assumption. Otherwise, if $ts' > md'.ts$, then by Lemma~\ref{la:sandwich} $ts' > md.ts$, and therefore $d_i$ replies with its local timestamp $ts$. According to the replica code, line~\ref{algreplica:setts} is the only place where $ts$ is changed. Furthermore, if $ts$ changes to $ts(wr')$ then $wr'$ is a \opwrite\ operation that committed. According to the \opwrite\ code, $wr'$ commits only after writing $ts(wr')$ to $\MDSD$. Hence, if $ts' > md'.ts$, then $rd$ invokes $\MDSD$ in line ~\ref{algreader:mdsdread2} and does so only after the corresponding \opwrite\ wrote $ts'$ to $\MDSD$. By Lemma~\ref{la:safety}, $\MDSD$ returns in line~\ref{algreader:mdsdread2} a value whose timestamp is a least $ts'$, which means that $md'.ts \geq ts'$, a contradiction.
\end{prooff}

\end{document}